\input phyzzx
\hsize=417pt %\vsize=600pt \baselineskip=20pt \maxdepth=0.2pt
\sequentialequations
\Pubnum={ TIT/HEP-334/COSMO-73, \hfill EDO-EP-5}
\date={ \hfill May 1996}
\titlepage
\vskip 32pt
\title{ Black Hole Singularity and Generalized Quantum Affine Parameter }
\author{ Akio Hosoya \footnote\dag {E-mail address: 
ahosoya@th.phys.titech.ac.jp}}
\vskip 12pt
\address{ Department of Physics, Tokyo Institute of Technology,
Oh-Okayama, Meguro-ku, Tokyo 152, JAPAN }                          
\vskip 15pt
\centerline{and}
\vskip 12pt
\author{Ichiro Oda \footnote\ddag {E-mail address: 
sjk13904@mgw.shijokyo.or.jp}}
\vskip 12pt
\address{ Edogawa University,                                
          474 Komaki, Nagareyama City,                        
          Chiba 270-01, JAPAN     }                          
%          
%
%              The titlepage ends at this place.
%
%======================================================================%
%
\abstract{ We study a behavior of quantum generalized affine parameter
(QGAP), which has been recently proposed by one of the present authors,
near the singularity and the event horizon in three and four spacetime
dimensions in terms of a minisuperspace model of quantum gravity.
It is shown that the QGAP is infinite to the singularity while it remains 
finite to the event horizon. This fact indicates a possible interpretation
that the singularity is wiped out in quantum gravity in this particular
model of black hole.} 
\endpage
%
%=========================================================================%
%
%        Macros
%

\def\sp(#1){\noalign{\vskip #1pt}}

%
%
%
%
%=========================================================================%
%
%   This part is the meat of body.
%
\topskip 30pt
\leftline{\bf 1. Introduction}
\par

One of the remarkable achievements in the classical theory of gravity
is surely the success of the proof of the well-known singularity
theorem by Hawking and Penrose [1]. The theorems show that under
rather physically reasonable assumptions spacetime singularities
are inevitable, according to the standard theory of general relativity, 
particularly both in the gravitational collapse of massive stars leading
to formation of black holes and in the past era in the big bang.

Even if the singularity theorems are so convincing to us at the present
time, since the time of the proof till this day many people have been
wondering whether such an approach based on the classical general
relativity makes sense to examine singularities at all without using 
quantum mechanics and/or quantum theory of gravity.

Intuitively a spacetime singularity is a ``place'' where gravitational fields 
become extremely large and curvature blows up, by which one would expect 
that classical theory breaks down anyway while quantum gravitational 
effects become so important. However, the difficulty lies in making the precise
definition of a singularity in quantum theory which is a natural 
generalization of geodesic incompleteness in classical relativity [2]. 
Thus it has been hoped that a union of quantum mechanics and 
general relativity would somehow avoid the formation of all types of 
singularities, and so would not encounter this difficulty.

Recently one of the present authors has proposed a criterion for a 
singularity to be quantum mechanically smeared away [3]. As a sort of
``order parameter'' measuring a physically meaningful length in  
quantum gravity, he has introduced a quantum mechanical version of 
Schmidt's generalization of classical affine parameter [4], 
which we call quantum generalized
 affine parameter (QGAP) in what follows. The QGAP is defined in four
spacetime  dimensions as
%%%%%%%%%%%%%%%%%%%%%%%%%%%Equation%%%%%%%%%%%%%%%%%%%%%%%%%%%%%%%%%%%
$$ \eqalign{ \sp(2.0)
\bigl|\alpha\bigr|_q = \int_{0}^{1}dt \sqrt{\sum_{a,b=0}^3 <
\bigl[V^{\mu}e^{b}_{\mu} \ \bigl( \ P \ \exp\int_0^{t} \omega \bigr)_b \ 
^a \bigr]^2 
>},
\cr
\sp(3.0)} \eqno(1)$$
%-----------------------------------------------------------------------
where $< \ >$ indicates an expectation value with respect to a quantum 
state, and $V^{\mu}$ is the tangent vector along a causal curve, which is 
not necessarily a geodesic curve. 
Incidentally, the classical generalized affine parameter (CGAP) is 
defined like the above but without the expectation value $< \ >$.
See the previous paper for more details [3]. The key idea behind this 
definition is that when one approaches a singularity the QGAP would 
become infinite owing to a large quantum fluctuation of the connection 
$\omega ^a \ _b$ although the CGAP remains finite. By using this 
definition it has been explicitly shown that this expectation is actually 
realized in the model of (2+1)-dimensional circular symmetric black hole 
[3]. At first sight, the definition (1) seems to suffer from the serious 
drawbacks that it manifestly depends on the initial vierbein as well as 
the initial point, and furthermore it never be gauge invariant, or 
BRST invariant, but this is an illusion. The reason is why this definition 
cannot distinguish two lengths differing by a finite value, but does 
the finiteness from the infiniteness, and hence gives a useful standard in 
measuring at least the difference between the finite and infinite Lorentz 
transformations.

However, in the retrospect we notice that the previous study [3] has some 
aspects which should be improved. One of them is that an explicit form of 
a quantum state is just assumed to be a gaussian wave packet spread over 
a mass a priori. This assumption is certainly too naive since the quantum 
state should be singled out from the constraint equations to the state 
a l$\acute a$ Dirac [5]. 
The other is the impossibility of applying the previous method for the 
more physically interesting (3+1)-dimensional black hole.

In this paper, we shall aim at overcoming these problems. The article 
is organized as follows. In section 2, we consider a minisuperspace model 
of (2+1)-dimensional black hole found by Ba$\tilde n$ados et al. [6]. We 
perform the canonical quantization of the model where mass and radial 
function are regarded as the canonical variables.  Here we shall quantize 
the black hole only in the interior region bounded by the singularity and 
the event horizon adopting the usual radial coordinate as a time 
parameter, and then solve the Wheeler-DeWitt equation in order to find 
the physical quantum state. By using such a selected quantum state, 
we show that the QGAP is infinite to the singularity  
while it is finite to the horizon. This fact indicates that the 
singularity is effectively infinitely far away and therefore the 
singularity is quantum mechanically wiped out by fluctuation of geometry. 
In section 3, the above analysis is extended to the 
case of the (3+1)-dimensional Schwarzschild black hole where a similar 
phenomenon can indeed occur. The last sectoin contains our conclusions. 
\vskip 12pt
\leftline{\bf 2. (2+1)-dimensional black hole}	
\par
We will begin by constructing a minisuperspace model of quantum gravity 
of the circular symmetric (2+1)-dimensinal black hole [6]. Let us 
consider the following spacetime metric
%%%%%%%%%%%%%%%%%%%%%%%%%%%%Equation%%%%%%%%%%%%%%%%%%%%%%%%%%%%%%%%%%%
$$ \eqalign{ \sp(2.0)
ds^2 &= - {1 \over {M(r) - {{\phi (r)}^2 \over l^2}}} dr^2 + ({M(r) - 
{{\phi (r)}^2 \over l^2}}) dt^2 + {\phi (r)}^2 d\theta^2 
\cr
     &= -(e^0)^2 + (e^1)^2 + (e^2)^2 . 
\cr
\sp(3.0)} \eqno(2)$$
%-----------------------------------------------------------------------
We have chosen the dreibeins as
%%%%%%%%%%%%%%%%%%%%%%%%%%%%Equation%%%%%%%%%%%%%%%%%%%%%%%%%%%%%%%%%%%
$$ \eqalign{ \sp(2.0)
e^0 &= {1 \over {\sqrt{M(r) - {{\phi (r)}^2 \over l^2}}}} dr\
     = {1 \over {\sqrt{f(r)}}} dr \  
\cr
e^1 &= \sqrt{M(r) -  {{\phi (r)}^2 \over l^2}} dt\
     = \sqrt{f(r)} dt
\cr
e^2 &= \phi (r) d \theta, \ 
\cr
\sp(3.0)} \eqno(3)$$
%-----------------------------------------------------------------------
where we have considered the mass of black hole $M(r)$ and the 
circumference radius 
$\phi (r)$ to be functions of only the radial coordinate $r$. Note that 
the negative cosmological constant $\Lambda$ is expressed by $\Lambda = - 
{1 \over l^2}$. Moreover, we have defined $f(r)$ to be $M(r) - {{\phi (r)}^2
 \over l^2}$ in this section for later convenience. The important point 
 is that the 
 radial coordinate $r$ plays a role of time since we would like to 
 perform a canonical quantization of the present model only inside the 
 event horizon $f(r)>0$.
 
 To make contact with the Einstein theory of gravity, let us impose the 
 torsion free condition in contrast with the 
 previous work [3] where the model having a torsion off-shell was 
 considered. From the torsion free equation, omitting the wedge 
 product,
%%%%%%%%%%%%%%%%%%%%%%%%%%%%Equation%%%%%%%%%%%%%%%%%%%%%%%%%%%%%%%%%%%
$$ \eqalign{ \sp(2.0)
0 = T^a := de^a + \omega^a \ _b  e^b,
\cr
\sp(3.0)} \eqno(4)$$
%-----------------------------------------------------------------------
it is straightforward to derive concrete expressions for the spin 
connection $\omega^a \ _b$ given by
%%%%%%%%%%%%%%%%%%%%%%%%%%%%Equation%%%%%%%%%%%%%%%%%%%%%%%%%%%%%%%%%%%
$$ \eqalign{ \sp(2.0)
\omega^0 \ _1 = \omega^1 \ _0 &=  {\dot f \over {2 \sqrt {f}}} e^1 \ 
\cr
\omega^0 \ _2 = \omega^2 \ _0 &= {{\dot \phi \sqrt{f} \over \phi}} e^2 \ 
\cr
otherwise &= 0,
\cr
\sp(3.0)} \eqno(5)$$
%-----------------------------------------------------------------------
where the dot denotes the differentiation with respect to $r$. Then the 
curvature 2-form is easily computed from the second Cartan structure 
equation. We obtain  
%%%%%%%%%%%%%%%%%%%%%%%%%%%%Equation%%%%%%%%%%%%%%%%%%%%%%%%%%%%%%%%%%%
$$ \eqalign{ \sp(2.0)
R^0 \ _1 &:= - R^2 = {1 \over 2} \ddot f e^0 e^1 
\cr
R^0 \ _2 &:= R^1 = {d \over {dr}}(\dot \phi \sqrt{f}) {\sqrt{f} \over \phi} e^0 e^2
\cr
R^1 \ _2 &:= R^2 = {{\dot f \dot \phi} \over {2 \phi}} e^1 e^2.
\cr
\sp(3.0)} \eqno(6)$$
%-----------------------------------------------------------------------
The classical action is now given by
%%%%%%%%%%%%%%%%%%%%%%%%%%%%Equation%%%%%%%%%%%%%%%%%%%%%%%%%%%%%%%%%%%
$$ \eqalign{ \sp(2.0)
S = -{2 \over {16 \pi G}} \int \ ( - e_a R^a + {1 \over l^2} e^0 e^1 e^2 ),
\cr
\sp(3.0)} \eqno(7)$$
%-----------------------------------------------------------------------
with $G$ being the Newton constant which is set to be 1 in what follows. 
By substituting Eqs.(3) and (6) into the right hand side of Eq.(7), this 
action can be written to be
%%%%%%%%%%%%%%%%%%%%%%%%%%%%Equation%%%%%%%%%%%%%%%%%%%%%%%%%%%%%%%%%%%
$$ \eqalign{ \sp(2.0)
S = \int \ dr \ L,
\cr
\sp(3.0)} \eqno(8)$$
%-----------------------------------------------------------------------
with the Lagrangian $L$ being 
%%%%%%%%%%%%%%%%%%%%%%%%%%%%Equation%%%%%%%%%%%%%%%%%%%%%%%%%%%%%%%%%%%
$$ \eqalign{ \sp(2.0)
L = T \ ( {1 \over 2} \dot f \dot \phi - {1 \over l^2} \phi ),
\cr
\sp(3.0)} \eqno(9)$$
%-----------------------------------------------------------------------
Here we have performed the integration with respect to $t$ with the time 
interval from $-2T$ to $2T$. By solving the Euler-Langrange 
equations derived from the action (9) we can check that the general solution 
essentially reproduces the solution $\phi = r$ and $M = constant$ 
which describes the 
(2+1)-dimensional black hole solution found by Ba$\tilde n$ados et al. [6]. 

Let us now carry out a canonical quantization of the present 
model. The canonical momenta $\pi_M$ and $\pi_{\phi}$ 
 conjugate to the canonical variables $M$ and $\phi$ respectively are 
given by 
%%%%%%%%%%%%%%%%%%%%%%%%%%%%Equation%%%%%%%%%%%%%%%%%%%%%%%%%%%%%%%%%%%
$$ \eqalign{ \sp(2.0)
\pi_M = {T  \over 2} \dot  \phi, \ 
\pi_{\phi} = T ( {1 \over 2} \dot M - {2 \over l^2} \phi \dot \phi).
\cr
\sp(3.0)} \eqno(10)$$
%-----------------------------------------------------------------------
Then the Hamiltonian becomes 
%%%%%%%%%%%%%%%%%%%%%%%%%%%%Equation%%%%%%%%%%%%%%%%%%%%%%%%%%%%%%%%%%%
$$ \eqalign{ \sp(2.0)
H = {T \over l^2} \phi + {2 \over T} \pi_{\phi} \pi_M + {4 \over T} {1 
\over l^2} 
\phi {\pi_M}^2,
\cr
\sp(3.0)} \eqno(11)$$
%-----------------------------------------------------------------------
which leads to the Wheeler-DeWitt(WDW) equation by the standard procedure 
in canonical gravity,
%%%%%%%%%%%%%%%%%%%%%%%%%%%%Equation%%%%%%%%%%%%%%%%%%%%%%%%%%%%%%%%%%%
$$ \eqalign{ \sp(2.0)
0 &= H \psi (M, \phi)
\cr
  &= \bigl( {T \over l^2} \phi - {2 \over T} {\partial^2 \over {\partial M 
  \partial \phi}}  -   {4 \over T}  {1 \over l^2} \phi {\partial^2 \over 
  \partial M^2} \bigr) \psi (M, \phi).
\cr
\sp(3.0)} \eqno(12)$$
%-----------------------------------------------------------------------
as the constraint to the state $\psi (M, \phi)$.
Note that as mentioned in the section 1, in the previous work [3] the 
Hamiltonian identically vanishes so that it is impossible to set up the 
physically meaningful WDW equation, however, this time we have the 
nontrivial WDW equation which is critical in selecting the physical state. 
Moreover, let us notice that the Hamiltonian (11) 
includes only 
the linear term with respect to $\pi_{\phi}$ which means that the WDW 
equation (12) has a common feature to the Schr$\ddot o$dinger equation 
in the configuration space [7]. 

Assuming the general solution of the WDW equation (12) to be the form 
$\psi(M, \phi) = \psi_1 (M) \psi_2 (\phi)$, we can solve the equation 
whose general solution is given by
%%%%%%%%%%%%%%%%%%%%%%%%%%%%Equation%%%%%%%%%%%%%%%%%%%%%%%%%%%%%%%%%%%
$$ \eqalign{ \sp(2.0)
\psi (M, \phi) = B \bigl(e^{\alpha M} - C e^{\beta M} \bigr) e^{{{T A } 
\over 4} \phi^2},
\cr
\sp(3.0)} \eqno(13)$$
%-----------------------------------------------------------------------
where $A$, $B$, and $C$ are integration constants, and $\alpha$ and 
$\beta$ are defined to be
%%%%%%%%%%%%%%%%%%%%%%%%%%%%Equation%%%%%%%%%%%%%%%%%%%%%%%%%%%%%%%%%%%
$$ \eqalign{ \sp(2.0)
\alpha &= - {{T A l^2} \over 8} - {T \over 2} \sqrt{1 + \bigl ({{A l^2} 
\over 4} \bigr )^2} ,
\cr
\beta  &= - {{T A l^2} \over 8} + {T \over 2} \sqrt{1 + \bigl ({{A l^2} 
\over 4} \bigr )^2} .
\cr
\sp(3.0)} \eqno(14)$$
%-----------------------------------------------------------------------

Now that we have a minisuperspace model of the (2+1)-dimensional quantum 
black hole, we can proceed to an evaluation of the quantum generalized 
affine parameter (QGAP). Before doing so, let us first consider the 
classical generalized affine parameter (CGAP) since this analysis brings 
us a clue leading to an important idea on attacking the QGAP later. 

Following the procedure adopted in the previous work [3], let us consider 
a curve along which $\footnote\dag {As stated in the previous paper [3] 
the curve should be specified without referring to the metric when we 
proceed to quantum gravity. So it would be more consistent if we replace 
Eq.(15) by the expression given in [3] which is written in terms of the 
coordinate only.}$ 
%%%%%%%%%%%%%%%%%%%%%%%%%%%%Equation%%%%%%%%%%%%%%%%%%%%%%%%%%%%%%%%%%%
$$ \eqalign{ \sp(2.0)
e^1 = 0, \ e^2 = k e^0, 
\cr
\sp(3.0)} \eqno(15)$$
%-----------------------------------------------------------------------
with $k$ being a constant other than zero. The value of $k$ determines 
the causal 
property of the path, that is, timelike for $|k|<1$, null for 
$|k|=1$, and spacelike for $|k|>1$. Now we would 
like to compute the CGAP from a point inside the horizon to the 
singularity. Physically we are interested in a situation where a causal 
(i.e. timelike or null) curve starting at the point $r_0$ inside 
the horizon at $t=0$ approaches the singularity $r=0$ at $t=1$. Thus 
it is sufficient to consider a curve near the singularity $r=0$. After 
taking the gauge $\phi = r$, let us define the following quantity for later 
convenience.
%%%%%%%%%%%%%%%%%%%%%%%%%%%%Equation%%%%%%%%%%%%%%%%%%%%%%%%%%%%%%%%%%%
$$ \eqalign{ \sp(2.0)
\gamma := \int \omega^2 \ _0 = \int {{\dot \phi \sqrt f} \over \phi} e^2 
= \int {{\dot \phi \sqrt f} \over \phi} k e^0 = k \log r , 
\cr
\sp(3.0)} \eqno(16)$$
%-----------------------------------------------------------------------
where we have used Eqs.(3), (5) and (15). By using this $\gamma$,  a 
straightforward calculation  gives CGAP as 
%%%%%%%%%%%%%%%%%%%%%%%%%%%%Equation%%%%%%%%%%%%%%%%%%%%%%%%%%%%%%%%%%%
$$ \eqalign{ \sp(2.0)
\bigl|\alpha\bigr|_c &:= \int_{0}^{1} dt \sqrt{\sum_{a,b=0}^2 
\bigl[V^{\mu} \ e^{b}_{\mu} \bigl( \ P \ \exp\int_0^t 
\omega \bigr)_b \ ^a \bigr]^2}
\cr
&= \int_{0}^{1} dt \sqrt{ 
\bigl[V^{\mu}e^{0}_{\mu} \ \bigl( \cosh \gamma - k \sinh \gamma , \ 0, 
- \sinh \gamma + k \cosh \gamma \bigr) \bigr]^2}
\cr
&\sim \int^0 dr {{1 + |k|} \over \sqrt {2 M}} r^{-|k|}. 
\cr
\sp(3.0)} \eqno(17)$$
%-----------------------------------------------------------------------
Thus it turned out that for a timelike curve $|k|<1$ the CGAP 
converges so that the path can reach the singularity $r=0$ in a finite 
length as physically expected, and on the other hand for a null $|k|=1$ or 
a spacelike path $|k|>1$ the CGAP diverges [3]. We can also 
calculate the CGAP from a point inside the event horizon to the event 
horizon in a perfectly similar way. Because $\gamma$ is finite at the 
horizon $r_H = l \sqrt M$ the CGAP becomes convergent for an arbitrary 
$k$ except $k=0$. 

Next let us consider the quantum generalized affine parameter (QGAP). The 
above classical analysis is very suggestive in the sense that the CGAP to 
the singularity is convergent for a timelike curve, and divergent for 
a null or spacelike curve. Perhaps in taking account of quantum effects 
the geometry would fluctuate violently particularly near the singularity 
so that a classically timelike curve may effectively become null or 
spacelike by which we have a possibility of having the divergent QGAP for 
an arbitrary curve. Later we will see that this attractive speculation is 
indeed the case except in the neighborhood of $k=0$.

In case of quantum theory, if we consider the same path (15) as in 
classical theory, by using Eq.(10) we can rewrite Eq.(16) to be
%%%%%%%%%%%%%%%%%%%%%%%%%%%%Equation%%%%%%%%%%%%%%%%%%%%%%%%%%%%%%%%%%%
$$ \eqalign{ \sp(2.0)
\gamma = {{2 k} \over T}  \int dr {1 \over r} \pi_M  = - i \ {{2 k} 
\over T} \int dr {1 \over r} {\partial \over \partial M}.
\cr
\sp(3.0)} \eqno(18)$$
%-----------------------------------------------------------------------
From Eqs.(1) and (17) the QGAP from a point inside the horizon to the 
singularity is given by 
%%%%%%%%%%%%%%%%%%%%%%%%%%%%Equation%%%%%%%%%%%%%%%%%%%%%%%%%%%%%%%%%%%
$$ \eqalign{ \sp(2.0)
\bigl|\alpha\bigr|_q =  \int_{0}^{1} dt \sqrt{< 
\bigl[V^{\mu}e^{0}_{\mu} \ \bigl( \cosh \gamma - k \sinh \gamma , \ 0, 
- \sinh \gamma + k \cosh \gamma \bigr) \bigr]^2>}
\cr
\sp(3.0)} \eqno(19)$$
%-----------------------------------------------------------------------
where $\gamma$ is now given by Eq.(18), and the expectation value $<F>$ 
for some operator $F$ is defined as
%%%%%%%%%%%%%%%%%%%%%%%%%%%%Equation%%%%%%%%%%%%%%%%%%%%%%%%%%%%%%%%%%%
$$ \eqalign{ \sp(2.0)
<\ F \ > :=  {1 \over {\int dM \psi^{\dagger}(M) \psi(M)}} 
\int dM \psi^{\dagger}(M) F \ \psi(M),
\cr
\sp(3.0)} \eqno(20)$$
%-----------------------------------------------------------------------
where $\psi(M)$ denotes the physical quantum state under the gauge $\phi 
= r$ satisfying the WDW equation (12). For simplicity, we shall confine 
ourselves to the case of the physical state (13) with $C=0$, since $C 
\not= 0$ case can be treated in a 
similar way without changing the essential results. 

As a first step, it is convenient to consider $< e^{ 2 \gamma} >$. 
From Eqs.(13) and (18) one obtains
%%%%%%%%%%%%%%%%%%%%%%%%%%%%Equation%%%%%%%%%%%%%%%%%%%%%%%%%%%%%%%%%%%
$$ \eqalign{ \sp(2.0)
e^{ 2 \gamma} \psi(M) = e^{ - i {{4 k} \over T} \alpha \log r } \psi(M) ,
\cr
\sp(3.0)} \eqno(21)$$
%-----------------------------------------------------------------------
where $\alpha$ is given by Eq.(14). Therefore from the definition of the 
expectation value (20) one gets
%%%%%%%%%%%%%%%%%%%%%%%%%%%%Equation%%%%%%%%%%%%%%%%%%%%%%%%%%%%%%%%%%%
$$ \eqalign{ \sp(2.0)
< e^{ 2 \gamma} > = e^{ - i {{4 k} \over T} \alpha \log r } .
\cr
\sp(3.0)} \eqno(22)$$
%-----------------------------------------------------------------------
To have non-zero value in Eq.(22) near $r \approx 0$, it is necessary 
that $\alpha$ should be a pure imaginary number according to the 
Riemann-Lebesgue lemma. Hence from Eq.(14) the integration constant $A$ 
needs to be expressed as $A = i a$ with $a$ being real number satisfying 
the inequality $|a| \geq {4 \over {l^2}}$.

Now it is easy to calculate the QGAP to the singularity $r=0$. The result 
is 
%%%%%%%%%%%%%%%%%%%%%%%%%%%%Equation%%%%%%%%%%%%%%%%%%%%%%%%%%%%%%%%%%%
$$ \eqalign{ \sp(2.0)
\bigl|\alpha\bigr|_q &\approx  \int^{0} dr {{\bigl|1 - {{|ka|} \over 
a}\bigr|} \over \sqrt{2 M}} \sqrt{ < e ^{ {{|ka|} \over {ka}} 2 \gamma} >}
\cr
&= \int^{0} dr {{\bigl|1 - {{|ka|} \over 
a}\bigr|} \over \sqrt{2 M}} r^{ - {{|ka|} \over 4} l^2 \bigl[ 1 + {|a|
\over a} \sqrt{ 1 - ( {4 \over {a l^2}})^2} \bigr]}.
\cr
\sp(3.0)} \eqno(23)$$
%-----------------------------------------------------------------------
It is obvious to see that the QGAP to the singularity becomes
convergent for the curve with 
$|k| <{ 4 \over { l^2 |a| \bigl[ 1 + {|a| \over a} \sqrt{ 1 - ( {4 
\over {a l^2}})^2} \bigr]}}$. On the other hand, it does divergent 
for the curve with $|k| \geq {4 \over { l^2 |a| \bigl[ 1 + {|a| \over a} 
\sqrt{ 1 - ( {4 \over {a l^2}})^2} \bigr]}}$.  In particular, 
when $a = -{4 \over l^2}$ we obtain  exactly the same result as the one 
in classical 
theory which suggests no quantum correction in this specific case. This 
can also be confirmed by checking the fact that the physical observable $\pi_M$ 
commuting with the Hamiltonian constraint (11) is equal 
in both classical and quantum 
theory just at this value. Now an important point is that for a 
sufficiently large $|a|$ the domain of $|k|$ of divergent QGAP near 
the singularity is much larger than the one of convergent QGAP . In the 
limit of $|a| \rightarrow \infty$, the QGAP to the singularity would 
become divergent for any curve except $k \not= 0$: the singularity is 
infinitely far away because of the strong quantum fluctuation of 
geometry. However, if we take the limit of $|a| \rightarrow \infty$, the 
physical state would approach zero. We consider a sufficiently large 
$|a|$ but not infinity to have a well-defined
physical state. Then it is 
concluded that the QGAP to the singularity is divergent for almost any 
path except in the neighborhood of $k = 0$ as mentioned before.

There remains a question of what becomes of paths having small $|k|$ near 
zero. Obviously as demonstrated in the above, if $|k|$ is small the QGAP is 
finite which means that $r=0$ is still a singularity according to the 
``classical'' definition of the singularity because one has at least 
one causal curve and thus is bundle-incomplete (b-incomplete) [1], [2], 
 [3], [4]. 
For this problem, the present authors shall take the following attitude. 
In classical relativity, the spacetime is said to be singular if at 
least one causal curve is b-incomplete. However, it seems to us that in 
quantum gravity ``at least one causal curve'' should be replaced by ``a 
generic curve'' since a single curve has a measure zero contribution to 
the path integral when we quantize the particle trajectory [3]. 

Here we would like to provide a circumstancial evidence which supports 
our argument.  First, Eqs.(3) and (15) give us the relation $k 
\approx {{\theta \sqrt M} \over {\log r}}$ near the singularity $r=0$ for 
the path (15). Thus by this equality the uncertainty $\Delta k$ is 
related to that of mass like ${{\Delta M} \over {\sqrt M \log r}}$ 
 and therefore, near
$r \approx 0$ the mass fluctuation $\Delta M$ would be  
huge provided that 
$\Delta k$ remains finite. 
This picture that the mass fluctuation near the singularity $r=0$
becomes large seems to be plausible from physical viewpoint.
In other words, we are not able to fix the value of $k$ to be a certain 
definite value owing to the quantum fluctuation

Next one expects the Lorentz boost to be so large near 
the classical singularity that the paths around the null trajectory 
($|k|=1$) would be most relevant to the contribution 
for the CGAP. Although the CGAP converges for such paths but $|k|=1$, 
the QGAP tends to diverge if taking a sufficiently large $|a|$. 

Before closing this section, let us make a comment on the QGAP to the event 
horizon from some point inside the horizon. Its evaluation is 
straightforward, and is finite for all paths when we take the same 
physical state adopted in the analysis of the QGAP to the singularity.
Thus the arguments done so far imply that in 
taking quantum effects into consideration the singularity is far away 
while the event horizon is not, hence the classical singularity is not 
the ``real'' singularity in quantum gravity.
\vskip 12pt
\leftline{\bf 3. (3+1)-dimensinal Schwarzschild black hole}	
\par
We now turn our attention to the more physically interesting Schwarzschild 
black hole in (3+1)-spacetime dimensions.
In this case we can follow  almost the same 
arguments as in the (2+1)-dimensional black hole. Since it is easily 
proved that the CGAP is finite both to the singularity and to the event 
horizon, let us concentrate on the analysis of the QGAP. The spacetime 
metric we now consider is
%%%%%%%%%%%%%%%%%%%%%%%%%%%%Equation%%%%%%%%%%%%%%%%%%%%%%%%%%%%%%%%%%%
$$ \eqalign{ \sp(2.0)
ds^2 &= -{1 \over { -1 + {{2 M(r)} \over \phi(r)}}} dr^2 + (-1 + {{2 
M(r)} \over \phi(r)}) dt^2 +  {\phi (r)}^2 ( d\theta^2 + \sin^2 
\theta d\varphi^2) 
\cr
     &= -(e^0)^2 + (e^1)^2 + (e^2)^2 +(e^3)^2, 
\cr
\sp(3.0)} \eqno(24)$$
%-----------------------------------------------------------------------
where the vierbeins take the form
%%%%%%%%%%%%%%%%%%%%%%%%%%%%Equation%%%%%%%%%%%%%%%%%%%%%%%%%%%%%%%%%%%
$$ \eqalign{ \sp(2.0)
e^0 &= {1 \over {\sqrt{-1 + {{2 M(r)} \over \phi(r)}}}} dr\
     = {1 \over {\sqrt{f(r)}}} dr \  
\cr
e^1 &= \sqrt{-1 + {{2 M(r)} \over \phi(r)}} dt\
     = \sqrt{f(r)} dt
\cr
e^2 &= \phi (r) d \theta, \ 
\cr
e^3 &= \phi (r) \sin \theta d \varphi,
\cr
\sp(3.0)} \eqno(25)$$
%-----------------------------------------------------------------------
where as in the (2+1)-dimensional case we have taken the dynamical 
variables to be the black hole mass $M(r)$ and the area radius 
$\phi(r)$. The variable $\phi(r)$ might correspond to the 
Teichm$\ddot u$ller parameter describing the ratio of the area of 
$S^2$ and the circumference of $S^1$ since the present spacetime has the 
topology $S^1 \times S^2$. In addition we have used the same symbol $f(r)$ to 
denote $-1 + {{2 M(r)} \over \phi(r)}$ now instead of $M(r) - {{{\phi(r)}^2} 
\over l^2}$ in three dimensions in the section 2.  
  
Then the torsion free condition (4) yields
%%%%%%%%%%%%%%%%%%%%%%%%%%%%Equation%%%%%%%%%%%%%%%%%%%%%%%%%%%%%%%%%%%
$$ \eqalign{ \sp(2.0)
\omega^0 \ _1 = \omega^1 \ _0 &= {\dot f \over {2 \sqrt {f}}} e^1 \ 
\cr
\omega^0 \ _2 = \omega^2 \ _0 &= {{\dot \phi \sqrt{f}} \over \phi} e^2 \ 
\cr
\omega^0 \ _3 = \omega^3 \ _0 &= {{\dot \phi \sqrt{f}} \over \phi} e^3 \
\cr
\omega^2 \ _3 = - \omega^3 \ _2 &= - {{\cot \theta} \over \phi} e^3 \ 
\cr
otherwise &= 0.
\cr
\sp(3.0)} \eqno(26)$$
%-----------------------------------------------------------------------
And the curvature 2-form is of the form
%%%%%%%%%%%%%%%%%%%%%%%%%%%%Equation%%%%%%%%%%%%%%%%%%%%%%%%%%%%%%%%%%%
$$ \eqalign{ \sp(2.0)
R^0 \ _1 &= {1 \over 2} \ddot f e^0 e^1 
\cr
R^0 \ _2 &= {d \over {dr}}(\dot \phi \sqrt{f}) {\sqrt{f} \over \phi} e^0 e^2
\cr
R^0 \ _3 &= {d \over {dr}}(\dot \phi \sqrt{f}) {\sqrt{f} \over \phi} e^0 e^3
\cr
R^1 \ _2 &= {{\dot f \dot \phi} \over {2 \phi}} e^1 e^2
\cr
R^1 \ _3 &= {{\dot f \dot \phi} \over {2 \phi}} e^1 e^3
\cr
R^2 \ _3 &= \bigl({1 \over{{\phi}^2}} + {{{\dot \phi}^2 f} \over{ 
\phi}^2} \bigr) e^2 e^3,
\cr
\sp(3.0)} \eqno(27)$$
%-----------------------------------------------------------------------
and the other components of $R^a \ _b$ vanish. By using these expressions 
the classical action, the Einstein-Hilbert action, given by
%%%%%%%%%%%%%%%%%%%%%%%%%%%%Equation%%%%%%%%%%%%%%%%%%%%%%%%%%%%%%%%%%%
$$ \eqalign{ \sp(2.0)
S = -{1 \over {16\pi G}} \int \ {1 \over 4} \varepsilon_{abcd} e^a e^b R^{cd} ,
\cr
\sp(3.0)} \eqno(28)$$
%-----------------------------------------------------------------------
can be recast to 
%%%%%%%%%%%%%%%%%%%%%%%%%%%%Equation%%%%%%%%%%%%%%%%%%%%%%%%%%%%%%%%%%%
$$ \eqalign{ \sp(2.0)
S &= \int \ dr \ L
\cr
  &= \int \ dr {T \over 4}  ( -1 + \dot f \dot \phi \phi  + {\dot \phi}^2 f ),
\cr
\sp(3.0)} \eqno(29)$$
%-----------------------------------------------------------------------
where $\varepsilon_{abcd}$ denotes the totally antisymmetric symbol with 
$\varepsilon_{0123} = +1$ and the integration with respect to $t$ has now 
done from $-{T \over 2}$ to ${T \over 2}$. The equations of motion derived 
from the reduced action (29) have in fact the solution $\phi = r$ and 
$M = constant$  
corresponding to the well-known Schwarzschild black hole. 

The canonical quantization is carried out by following the procedure in 
the last section. The canonical conjugate momenta now have the form
%%%%%%%%%%%%%%%%%%%%%%%%%%%%Equation%%%%%%%%%%%%%%%%%%%%%%%%%%%%%%%%%%%
$$ \eqalign{ \sp(2.0)
\pi_M = {T  \over 2} \dot \phi, \ 
\pi_{\phi} = T ( {1 \over 2} \dot M - {1 \over 2} \dot \phi),
\cr
\sp(3.0)} \eqno(30)$$
%-----------------------------------------------------------------------
and the Hamiltonian is of the form 
%%%%%%%%%%%%%%%%%%%%%%%%%%%%Equation%%%%%%%%%%%%%%%%%%%%%%%%%%%%%%%%%%%
$$ \eqalign{ \sp(2.0)
H = {T \over 4} + {2 \over T} \pi_{\phi} \pi_M + {1 \over T} {\pi_M}^2,
\cr
\sp(3.0)} \eqno(31)$$
%-----------------------------------------------------------------------
which leads to the Wheeler-DeWitt(WDW) equation given by
%%%%%%%%%%%%%%%%%%%%%%%%%%%%Equation%%%%%%%%%%%%%%%%%%%%%%%%%%%%%%%%%%%
$$ \eqalign{ \sp(2.0)
0 &= H \psi (M, \phi)
\cr
  &= \bigl({T \over 4} - {2 \over T} {{\partial^2} \over {\partial M \partial 
  \phi}} -   {1 \over T} {{\partial^2} \over {\partial M^2}} \bigr) 
  \psi (M, \phi).
\cr
\sp(3.0)} \eqno(32)$$
%-----------------------------------------------------------------------
The general solutions for the WDW equation (32) can be searched by the 
method of separation of variables  
as done in the previous section. From the method, the following 
solution can be obtained 
%%%%%%%%%%%%%%%%%%%%%%%%%%%%Equation%%%%%%%%%%%%%%%%%%%%%%%%%%%%%%%%%%%
$$ \eqalign{ \sp(2.0)
\psi (M, \phi) = B \bigl(e^{\alpha M} - C e^{\beta M}\bigr) e^{{{T A } 
\over 2} \phi},
\cr
\sp(3.0)} \eqno(33)$$
%-----------------------------------------------------------------------
with $A$, $B$, and $C$ being integration constants, and $\alpha$ and 
$\beta$ are now defined by
%%%%%%%%%%%%%%%%%%%%%%%%%%%%Equation%%%%%%%%%%%%%%%%%%%%%%%%%%%%%%%%%%%
$$ \eqalign{ \sp(2.0)
\alpha &= - {{T A} \over 2} - {T \over 2} \sqrt{1 + A^2 } ,
\cr
\beta  &= - {{T A} \over 2} + {T \over 2} \sqrt{1 + A^2 } .
\cr
\sp(3.0)} \eqno(34)$$
%-----------------------------------------------------------------------

Now we can go on to an evaluation of the QGAP in four dimensions. Instead 
of the curve (15) let us consider the following curve
%%%%%%%%%%%%%%%%%%%%%%%%%%%%Equation%%%%%%%%%%%%%%%%%%%%%%%%%%%%%%%%%%%
$$ \eqalign{ \sp(2.0)
e^1 = k e^0, \  e^2 = e^3 = 0,
\cr
\sp(3.0)} \eqno(35)$$
%-----------------------------------------------------------------------
whose causal property is determined by $k$ as in three dimensions. Under the 
gauge choice $\phi=r$, this time $\gamma$ is defined as
%%%%%%%%%%%%%%%%%%%%%%%%%%%%Equation%%%%%%%%%%%%%%%%%%%%%%%%%%%%%%%%%%%
$$ \eqalign{ \sp(2.0)
\gamma &:= \int \omega^1 \ _0 = \int {{\dot f} \over {2 \sqrt f}} e^1 
= \int {{\dot f} \over {2 \sqrt f}} k e^0 
\cr
&= - {k \over T}  \int dr {1 
\over r} 
\pi_M -  {{k T} \over 4} \int dr {1 \over {2M - r}} {\pi_M}^{-1},
\cr
\sp(3.0)} \eqno(36)$$
%-----------------------------------------------------------------------
where the Hamiltonian constraint $H \approx 0$ was used in deriving the 
last equation. Again for simplicity, we restrict our consideration  
to the physical state (33) with $C=0$ without loss of generality. After setting 
$A=i \ a \ (a \in \bf R)$, from Eqs.(33) and (36), one obtains 
%%%%%%%%%%%%%%%%%%%%%%%%%%%%Equation%%%%%%%%%%%%%%%%%%%%%%%%%%%%%%%%%%%
$$ \eqalign{ \sp(2.0)
e^{ 2 \gamma} \psi(M) = e^{i {{2 k} \over T} \alpha \log r + i {{k T} \over 
{2 \alpha}} \log(2M - r)} \psi(M) ,
\cr
\sp(3.0)} \eqno(37)$$
%-----------------------------------------------------------------------
where $\alpha$ is given by 
%%%%%%%%%%%%%%%%%%%%%%%%%%%%Equation%%%%%%%%%%%%%%%%%%%%%%%%%%%%%%%%%%%
$$ \eqalign{ \sp(2.0)
\alpha = - i  {{a + \sqrt{a^2 - 1}} \over 2}  T  ,
\cr
\sp(3.0)} \eqno(38)$$
%-----------------------------------------------------------------------
with $|a| \geq 1$. At this stage, we remark on a subtlety which does not 
exist in comparison with 
(2+1)-dimensional balck hole. Namely in evaluating (37), we have 
nontrivial commutation relations between $2M$ and $\pi_M$ or 
$\pi_{M}^{-1}$, which give rise to additional terms  in the 
right hand side of Eq.(37). In this article we have neglected their 
contributions. More delicate evaluation of them will be certainly needed.
Thus from Eqs.(20), (37) and (38) the expectation 
value of $e^{2 \gamma}$ can be obtained as follows:
%%%%%%%%%%%%%%%%%%%%%%%%%%%%Equation%%%%%%%%%%%%%%%%%%%%%%%%%%%%%%%%%%%
$$ \eqalign{ \sp(2.0)
< e^{ 2 \gamma} > = e^{{ k (a + \sqrt{a^2 - 1}) \log r }} {1 \over {\int 
dM}} 
\int dM e^{- k (a - \sqrt{a^2 - 1}) \log(2M - r) }.
\cr
\sp(3.0)} \eqno(39)$$
%-----------------------------------------------------------------------
Here it is necessary to define the inner product more precisely. Since we 
are taking account of only the interior region of the event horizon of 
the black hole, we shall take the integration region from $r \over 2$ to the 
cutoff of large mass $M_0$ in order to keep the norm finite. Under this 
definition of the inner product, it is easy to calculate the last 
integration in the right hand side of Eq.(39). The result is  
%%%%%%%%%%%%%%%%%%%%%%%%%%%%Equation%%%%%%%%%%%%%%%%%%%%%%%%%%%%%%%%%%%
$$ \eqalign{ \sp(2.0)
\int_{r \over 2}^{M_0} dM e^{- k (a - \sqrt{a^2 - 1}) \log(2M - r)} 
= {1 \over 2} \ {1 \over { 1 - k (a - \sqrt{a^2 - 1})}} x^{ 1 -  k (a - 
\sqrt{a^2 - 1})} \bigg|_{x=0}^{x=2M_0},
\cr
\sp(3.0)} \eqno(40)$$
%-----------------------------------------------------------------------
where we have performed a change of variable $x = 2M - r$. The important 
point is that this integration is finite at $x=0$, i.e., $r=2M$ when $k$ 
satisfies the inequality 
%%%%%%%%%%%%%%%%%%%%%%%%%%%%Equation%%%%%%%%%%%%%%%%%%%%%%%%%%%%%%%%%%%
$$ \eqalign{ \sp(2.0)
 k (a - \sqrt{a^2 - 1}) \leq 1.
\cr
\sp(3.0)} \eqno(41)$$
%-----------------------------------------------------------------------
This inequality holds for almost all positive $k$ if $|a|$ is sufficiently 
large.
The sign of $k$ has a simple physical interpretation. Imagine a particle  
approaching the event horizon or the singularity along the path (35). 
Substituting (25) into (35), one sees that $k$ is positive for a 
path approaching the event horizon, on the other hand, negative for the 
one to the singularity. For the time being let us assume that the 
inequality (41) is satisfied .

Now we would like to examine the behavior of the QGAP to the singularity. 
To do so, it is sufficient to consider only the potentially divergent 
part of $< e^{2 \gamma} >$. We set 
%%%%%%%%%%%%%%%%%%%%%%%%%%%%Equation%%%%%%%%%%%%%%%%%%%%%%%%%%%%%%%%%%%
$$ \eqalign{ \sp(2.0)
< e^{ 2 \gamma} > \sim e^{ k (a + \sqrt{a^2 - 1}) \log r } .
\cr
\sp(3.0)} \eqno(42)$$
%-----------------------------------------------------------------------
Then the QGAP to the singularity can be calculated in a perfectly similar 
manner to the case of 2+1 dimensional black hole. We obtain
%%%%%%%%%%%%%%%%%%%%%%%%%%%%Equation%%%%%%%%%%%%%%%%%%%%%%%%%%%%%%%%%%%
$$ \eqalign{ \sp(2.0)
\bigl|\alpha\bigr|_q &\approx   \int_0^1 dt \bigl|{dr \over dt}\bigr| {1 \over 
\sqrt{f}} 
{{\bigl|1 - {{|a|} \over 
a} k \bigr|} \over \sqrt{2}} \sqrt{ < e ^{ {{|a|} \over {a}} 2 \gamma} >}
\cr
&\propto \int^{0} dr \sqrt{r} \ r^{  {1 \over 2} k |a| ( 1 + 
{{|a|} \over {a}} \sqrt { 1 - {1 \over a^2} } \ )}
\cr
&= {1 \over { {1 \over 2} \bigl[ 3 + k |a| ( 1 + {{|a|} \over {a}}
\sqrt { 1 - {1 \over a^2} } \ ) } \bigr]}
r^{  {1 \over 2} \bigl[ 3 + k |a| ( 1 + {{|a|} \over {a}}
\sqrt { 1 - {1 \over a^2} } \ )  \bigr] } \bigg|^0.
\cr
\sp(3.0)} \eqno(43)$$
%-----------------------------------------------------------------------
Therefore, the QGAP to the singularity diverges if $3 + k |a| ( 1 + {{|a|}
 \over {a}}
\sqrt { 1 - {1 \over a^2} } \ )< 0$ which is satisfied for almost all negative 
$k$ when $|a|$ is sufficiently large. Moreover, it is easy to 
show that by using the same physical state the QGAP to the event horizon 
converges when $|a|$ is so large. At this point, it is remarkable to 
notice that the QGAP is infinite to the singularity while the QGAP is 
strictly finite to the event horizon by taking the same quantum state
with very large $|a|$. In this way, we have shown that 
the ``classical'' singularity $r=0$ in the Schwarzschild black hole is 
not a singularity in quantum gravity in the sense of the conventional 
definition of a singularity [2], [3], [4].
\vskip 12pt
\leftline{\bf 4. Conclusion}	
\par
In this article, we have shown that the singularities in both three 
dimensional BTZ balck hole [6] and four dimensional Schwarzschild black 
hole can be smeared and is infinitely far away if we take an effect of 
quantum gravity in the specific minisuperspace models. Of course, 
we do not intend to claim that we have proven a smearing, as a result,
a disappearance of the 
classical singularities in quantum gravity. However, our model seems to 
reflect essential characteristic features of quantum black holes so 
that our investigation strongly suggests that the present analysis can be 
generalized to other minisuperspace models and even to full quantum 
gravity.

For further development, it would be very interesting to couple various 
matter fields to the present model and to apply for a proof of the 
strong cosmic censorship which is now under investigation. And recently, 
Horowitz and Marolf [8] discussed a possibility of self-adjoint extension 
of the Laplacian operator in curved spacetimes to a classically singular 
point at which geodesic is incomplete. If the extension is possible, the 
singularity is smeared out in the context of quantum field theory in 
curved spacetime. It might be also interesting to examine the relation 
between our formalism and theirs.

\vskip 12pt
\leftline{\bf References}
\centerline{ } %
\par
\item{[1]} S.W.Hawking and R.Penrose, Proc. Roy. Soc. Lond., 
{\bf A314} (1970) 529. 

\item{[2]} R.M.Wald, General Relativity (The University of Chicago Press, 1984);
 F.J.Tipler, C.J.S.Clarke and G.F.R.Ellis, in General Relativity, ed. 
A.Held, vol.2 (Plenum Press, 1980), p.97

\item{[3]} A.Hosoya, Class. Quant. Grav.12 (1995) 2967. 

\item{[4]} B.G.Schmidt, Gen. Rel. and Grav.{\bf 1} (1971) 269; S.Hawking 
and G.F.R.Ellis, The large scale structure of space-time (Cambridge U.P., 
Cambrigde, 1973), Chap.8.

\item{[5]} P.A.M.Dirac, Lectures on Quantum Mechanics (Yeshiva University, 1964)

\item{[6]} M.Ba$\tilde n$ados, C.Teitelboim and J.Zanneli, Phy. Rev. Lett. 
{\bf 69} (1992) 1849.

\item{[7]} A.Ashtekar, Lectures on Non-perturbative Canonical Gravity (World 
Scientific Publishing, 1991), Chap.12

\item{[8]} G.Horowitz and D.Marolf, Phy. Rev. {\bf D52} (1995) 5670. 
{\bf 69} (1992) 1849.

\endpage
%

%=======================================================================%
%
\bye